\documentclass[twocolumn,prl]{revtex4}

\usepackage[latin1]{inputenc}
\usepackage[T1]{fontenc}
\usepackage{lmodern}
\usepackage{eurosym}
\usepackage[frenchb, english]{babel}
\usepackage{dsfont,color,units,pstricks}
\usepackage{amsfonts}
\usepackage{amsmath}
\usepackage{amsthm}
\usepackage{amssymb}
\usepackage{varioref}
\usepackage{listings}
\usepackage{graphicx}
\usepackage{float}
\usepackage{color}

\newcommand \be  {\begin{equation}}
\newcommand \beno  {\begin{equation*}}
\newcommand \bea {\begin{eqnarray} \nonumber }
\newcommand \ee  {\end{equation}}
\newcommand \eeno  {\end{equation*}}
\newcommand \eea {\end{eqnarray}}
\newcommand{\Tr}{\text{Tr}}

\begin{document}

\title{Instanton Approach to Large $N$ Harish-Chandra-Itzykson-Zuber Integrals}

\author{J. Bun$^{1,2,3,\footnote{joel.bun@u-psud.fr}}$, J. P. Bouchaud$^1$, S. N. Majumdar$^2$, M. Potters$^1$}
\affiliation{$^1$ Capital Fund Management, 23--25, rue de l'Universit\'e, 75\,007 Paris, France}
\affiliation{$^2$ CNRS, LPTMS, Batiment 100, Universit\'e d'Orsay, 91405 Orsay Cedex, France}
\affiliation{$^3$ DeVinci Finance Lab, P{\^o}le Universitaire L{\'e}onard de Vinci, 92916 Paris La D{\'e}fense, France}

\begin{abstract} 
We reconsider the large $N$ asymptotics of Harish-Chandra-Itzykson-Zuber integrals. We provide, using Dyson's Brownian motion and the method of instantons, 
an alternative, transparent derivation of the Matytsin formalism for the unitary case. Our method is easily generalized to 
the orthogonal and symplectic ensembles. We obtain an explicit solution of Matytsin's equations in the case of Wigner matrices, as well as a general expansion 
method in the dilute limit, when the spectrum of eigenvalues spreads over very wide regions.
\end{abstract}

\sloppy
\maketitle


The ability to perform explicit calculations of sums and integrals is at the heart of much groundbreaking progress in
theoretical physics, in particular, in field theory or statistical mechanics. In that respect, the so-called Harish-Chandra-Itzykson-Zuber (HCIZ)
integral \cite{harish1957differential,itzykson1980planar} is among the most beautiful results, and has found several applications in many different fields, including Random Matrix Theory, disordered
systems or quantum gravity (for a particularly insightful introduction, see \cite{tao-blog}). The generalized HCIZ integral ${\cal I}_\beta(A,B)$ is defined as:
\be\label{HCIZ-def}
{\cal I}_\beta(A,B) = \int_{G(N)} {\cal D} \Omega \,\, e^{\frac{\beta N}{2} \Tr  A \Omega B \Omega^{\dag}},
\ee
where the integral is over the (flat) Haar measure of the compact group $\Omega \in G(N)=O(N), U(N)$ or $Sp(N)$ in $N$ dimensions and $A,B$ are arbitrary 
$N \times N$ symmetric (hermitian or symplectic) matrices. The parameter $\beta$ is the usual Dyson ``inverse temperature'', with $\beta=1,2,$ or $4$, respectively for the three groups. 
In the unitary case 
$G(N)=U(N)$ and $\beta=2$, it turns out that the HCIZ integral can be expressed exactly, for all $N$, as the ratio of determinants that depend on $A,B$, and additional $N$-dependent prefactors:
\be
{\cal I}_{\beta=2}(A,B)=\frac{c_{N}}{N^{(N^2 - N)/2}} \frac{\det\left( (e^{N \nu_{i} \lambda_{j}})_{1 \le i,j \le N}\right)}{\Delta(A) \Delta(B)}
\ee
with $\{\nu_i\}$, $\{\lambda_i\}$ the eigenvalues of $A$ and $B$, $\Delta(A) = \prod_{i<j} |\nu_{i} - \nu_{j}|$ the Vandermonde determinant of $A$ 
[and, similarly, for $\Delta(B)$], and $c_{N} = \prod_{i}^{N} i!$. 

Although the HCIZ result is fully explicit for $\beta=2$, the expression in terms of determinants is highly nontrivial 
and quite tricky. For example, the expression becomes degenerate ($0/0$) whenever two
eigenvalues of $A$ (or $B$) coincide. Also, as is well known, determinants contain $N!$ terms of alternating signs, which makes their order of magnitude 
very hard to estimate {\it a priori}. This difficulty appears clearly when one is interested in the large $N$ asymptotics of HCIZ integrals, for which one 
would naively expect to have a simplified, explicit expression as a functional $F_2(\rho_A,\rho_B) = \lim_{N \to \infty} N^{-2} \ln {\cal I}_{\beta=2}(A,B)$ 
of the eigenvalue densities $\rho_{A,B}$ of $A,B$. [The $N^{-2}$ scaling can be guessed by noting that generically $\Tr  A \Omega B \Omega^{\dag} = O(N)$, but of course this
is insufficient]. But even this large $N$ limit turns out to be highly nontrivial. In a remarkable paper, 
Matytsin \cite{matytsin1994large} suggested a mapping to a nonlinear hydrodynamical problem in one-dimension, the solution of which gives, in principle,
access to $F_2(\rho_A,\rho_B)$. Matytsin's result for $N \to \infty$ was later shown by Guionnet and Zeitouni \cite{guionnet2002large} to be mathematically rigorous. Still, neither Matytsin's, nor 
Guionnet and Zeitouni's derivation is very transparent (at least to our eyes). In this Letter, we recover Matytsin's equations using a rather straightforward instanton approach 
to the large deviations of the Dyson Brownian motion that describes the (fictitious) dynamics of eigenvalues connecting $\rho_A$ to $\rho_B$. Our approach is easily adapted to {\it arbitrary values} of $\beta$, including 
the orthogonal case which yields Zuber's ``$\frac12$-rule'' when $N \to \infty$, i.e. $F_1(\rho_A,\rho_B)=\, F_2(\rho_A,\rho_B)/2$ \cite{zuber2008large}. 
We then solve exactly Matytsin's equation in two particular cases (i) both $\rho_A$ and $\rho_B$ are Wigner semicircle distributions (of arbitrary widths $\sigma_{A,B}$); (ii) $\rho_A$ and $\rho_B$ 
are arbitrary, but with diverging widths $\sigma_{A,B} \to \infty$. We compare our results with the small-$\sigma$ expansion obtained in \cite{collins2003moments}. 

\begin{figure}
 \centering
   \includegraphics[width=1\columnwidth]{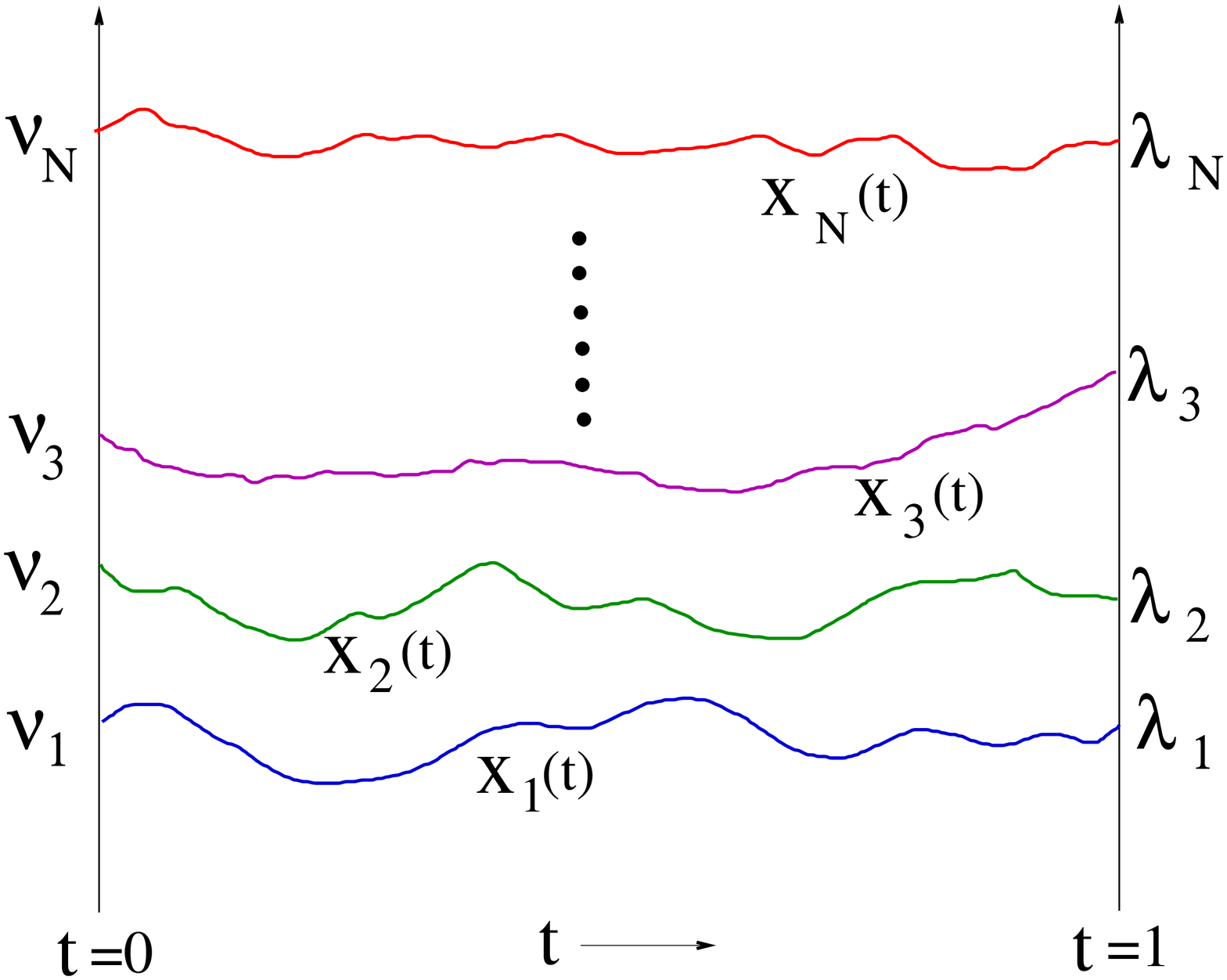}
 \caption{Dyson Brownian motion transporting the initial distribution $\rho_A$ of the eigenvalues of $A$ to the final distribution $\rho_B$ of the
 eigenvalues of $B$, in a (fictitious) time $t=1$.}
  \label{fig:dyson}
\end{figure}

Our main idea is to study, using the method of instantons, the large deviations of the Dyson Brownian motion of 
eigenvalues that brings an initial distribution of eigenvalues $\rho_A$ to a final distribution $\rho_B$ (see Fig. 1). This occurs with a probability that is exponentially small, $\propto \exp(-N^2 S)$, 
with a rate $S$ that we are able to relate directly to the HCIZ integral -- see below. (The idea to use Dyson Brownian motion in that context can also be found,
but in a very different language, in \cite{guionnet2004}.) Suppose that one adds to a certain matrix $A$ small random Gaussian Hermitian matrices of variance ${\rm d}t/N$. 
It is well known that in the ${\rm d}t \to 0$ limit, the eigenvalues  $\{ x_i \}$ of the time-dependent matrix evolve according to (see \cite{dyson1962brownian}): 
\be \label{Dyson}
{\rm d}x_i = \sqrt{\frac{2}{\beta N}} {\rm d}W + \frac{1}{N} {\rm d}t \sum_{j \neq i} \frac{1}{x_i - x_j},
\ee
where $W(t)$ is the standard Brownian motion and we set henceforth $\beta \equiv 2$, corresponding to unitary matrices. The calculation of $S$ can be done using two different (but complementary) languages: 
that of particle trajectories and that of densities, using the Dean-Kawasaki formalism. We start with the particle point of view, and sketch the density functional 
method later. We introduce the total potential energy 
$U\left(\{ x_i \}\right) = - \frac{1}{{N}} \sum_{i < j} \ln | x_i - x_j |$, and the corresponding ``force'' $f_i = - \partial_{x_i} U$. 
The probability of a given trajectory for the $N$ Brownian motions between time $t=0$ and time $t=1$ is given by (see Fig. 1): \footnote{We neglect the Jacobian which is small 
in the large $N$ (small temperature) limit, as usual.}
\bea \nonumber
{\cal P}(\{ x_i(t) \}) &=& {\cal N} \exp - \left[ \frac{N}{2} \int_0^1 {\rm d} t \sum_i \left( \dot x_i + \partial_{x_i} U \right)^2 \right ] \\
&\equiv&  {\cal N} e^{-N^2 S},
\eea
where ${\cal N}$ is some normalization. The action $S=S_1+S_2$ contains a total derivative equal, in the continuum limit, to:
\be
\label{action_VdM}
S_1=-\frac{1}{2} \left[\int {\rm d}x {\rm d}y \rho_Z(x) \rho_Z(y) \ln |x - y| \right]^{Z=B}_{Z=A} 
\ee
and:
\be \label{action}
S_2=\frac{1}{2N}  \int_0^1 {\rm d}t \sum_{i=1}^N  \left[\dot x_i^2 + (\partial_{x_i} U)^2 \right]
\ee
The ``instanton'' trajectory that dominates the probability for large $N$ is such that the functional derivative with respect to all $x_i(t)$ is zero 
(see e.g. \cite{bray}):
\be \label{motion}
-2 \frac{d^2 x_i}{dt^2} + 2 \sum_{\ell=1}^N \partial^2_{x_i,x_\ell} U  \partial_{x_\ell} U= 0
\ee
which leads, after a few algebraic manipulations, to
\be
\frac{d^2 x_i}{dt^2} = - \frac{2}{N^2} \sum_{\ell \neq i} \frac{1}{(x_i - x_\ell)^3}.
\ee
This can be interpreted as the motion of unit mass particles, accelerated by an {\it attractive} force that derives from an effective two-body potential $\phi(r)= -(Nr)^{-2}$. 
The hydrodynamical description of such a fluid is given by the Euler equations for the density $\rho(x,t)$ and the velocity field $v(x,t)$ \footnote{Actually,
the Dean-Kawasaki formalism allows one to see that a viscosity term, of order $N^{-1}$, is in fact also present -- see below.}
\be\label{NS1}
\partial_t \rho(x,t) + \partial_x [\rho(x,t)v(x,t)] = 0,
\ee
and
\be\label{NS2}
\partial_t v(x,t) + v(x,t) \partial_x v(x,t) = - \frac{1}{\rho(x,t)} \partial_x P(x,t),
\ee
where $P(x,t)$ is the pressure field which reads, from the virial formula in one dimension  \cite[p. 138]{virial}:
\be
P = \rho T  - \frac{1}{2} \rho \sum_{\ell \neq i} |x_i-x_\ell| \phi'(x_i - x_\ell) \approx - \frac{\rho}{N^2} \sum_{\ell \neq i} \frac{1}{(x_i - x_\ell)^2},
\ee
because the fluid is at an effective temperature $T=1/N$ (see below). Now, using the same argument as Matytsin \cite{matytsin1994large}, i.e, 
writing $x_i - x_\ell \approx (i - \ell)/(N\rho)$  and $\sum_{n=1}^\infty n^{-2} = \frac{\pi^2}{6}$, one finally finds \footnote{This also implicitely assumes 
that assuming that the density $\rho$ vanishes at the 
edges of the spectrum.}
\be
P(x,t) = - \frac{\pi^2}{3} \rho(x,t)^3,
\ee
and therefore Matytsin's equations for $\rho$ and $v$. Plugging this back in the action $S$, and going to the continuous limit, one also finds:
\be
\label{action_instanton}
S_2 \approx \frac{1}{2} \int_{0}^{1}dt \int {\rm d} x \rho(x,t) \left[v^2(x,t) + \frac{\pi^2}{3} \rho^2(x,t) \right],
\ee
which is exactly Matytsin's action \cite{matytsin1994large}. Finally, the probability ${\cal P}(\{ \lambda_i \} | \{ \nu_i \})$ to observe the set of eigenvalues $\{\lambda_i\}$ 
of $B$ for a given set of eigenvalues $\nu_i$ for $A$ is proportional to $\exp[-N^2(S_1+S_2)]$ where $S_2$ is obtained by plugging into Eq. (\ref{action_instanton}) 
the solution of the Euler equations (\ref{NS1}, \ref{NS2}), with $v(x,t=\{0,1\})$ chosen in such a way that $\rho(\nu,t=0)=\rho_A(\nu)$ and $\rho(\lambda,t=1)=\rho_B(\lambda)$.

Now, the idea is to interpret the HCIZ integrand in the unitary case, $\exp[N \Tr  A U B U^{\dag}]$, as a part of the propagator of the diffusion operator in the space of Hermitian matrices. 
Indeed, adding to $A$ small random Gaussian Hermitian matrices of variance ${\rm d}t/N$, the probability to end up with matrix $B$ in a time $t=1$ is 
${\cal P}(B|A) \propto \exp -[N/2\,\, \Tr(A - B)^2]$. Writing $B = V \Lambda V^{\dag}$ with $\Lambda={\mbox{diag}}(\lambda_1,
\dots, \lambda_N)$, the change of variables, as is well known, induces a probability measure on $\{\lambda_i\}$ alone that includes a Vandermonde determinant 
$\Delta^2(B)=\prod_{i<j} |\lambda_i - \lambda_j|^2$. Since the conditional distribution of $\{ \lambda_i \}$ is obviously invariant under $B \to U B U^{\dag}$ 
where $U$ is an arbitrary unitary transformation, we get another expression for ${\cal P}(\{ \lambda_i \} | \{ \nu_i \})$ \footnote{A more rigorous derivation in the unitary case $\beta=2$, that
includes all prefactors, uses Johansson's formula \cite{Johansson}.}:
\bea \label{identification}
& &{\cal P}(\{ \lambda_i \} | \{ \nu_i \}) 
\nonumber \\
&\propto& \prod_{i<j} |\lambda_i - \lambda_j|^2 \int {\cal D}U  \exp -[\frac{N}{2} \Tr(A - U B U^{\dag})^2] \nonumber \\
&\propto& \Delta^2(B) \exp -[\frac{N}{2}(\Tr \, A^2 + \Tr \, B^2)] \,\, {\cal I}_{2}(A,B).
\eea
Comparing this last expression for $\beta=2$ with the above calculation, and taking care of the proportionality coefficients, we get as a final expression for 
$F_{\beta=2}(A,B) = \lim_{N \to \infty} N^{-2} \ln \, {\cal I}_2(A,B)$:
\bea \label{final} \nonumber
& &F_{2}(A,B) = -\frac34-S_2(A,B) + \frac{1}{2} \int {\rm d}x\, x^2 (\rho_A(x)+\rho_B(x))\\ 
& & - \frac{1}{2} \int {\rm d}x {\rm d}y\, [\rho_A(x) \rho_A(y)+\rho_B(x) \rho_B(y)] \ln |x - y|, 
\eea
which is, apart from the $-3/4$ term which comes from the prefactor in Eq. (2), precisely Matytsin's result \cite{matytsin1994large}. Now, the whole calculation above can be repeated for the $\beta=1$ (orthogonal group) or
$\beta=4$ (symplectic group) with the final (simple) result $F_{\beta}(A,B)=\beta F_{2}(A,B)/2$. This coincides with the result obtained by Zuber in the orthogonal case $\beta=1$ \cite{zuber2008large} 
(see also \cite{guionnet2004,collins2009asymptotics}).

We now briefly explain how to obtain the same result using the Dean-Kawasaki framework \cite{kawasaki,dean1996langevin}. As shown by Dean \cite{dean1996langevin}, the density $\rho(x,t)$ of interacting particles obeying the Langevin equation (\ref{Dyson}) 
is found to satisfy the (functional) Langevin equation $\partial_t \rho(x,t) + \partial_x J(x,t) = 0$, with
\bea \nonumber
J(x,t) &=& \frac{1}{N}\xi(x,t) \sqrt{\rho(x,t)} -  \frac{1}{2N} \partial_x \rho(x,t) \\
&-&  \rho(x,t) \int {\rm d}y \partial_x V(x-y) \rho(y,t),
\eea
where $V(r)=- \ln r$ is the two-body interaction potential, $\xi(x,t)$ is a normalized Gaussian white noise (in time and in space), and, unlike in \cite{dean1996langevin}, we define $\rho(x,t) = \frac{1}{N} \sum_{i=1}^N \delta[x-x_i(t)]$.
One can again write the weight of histories of $\{ \rho(x,t) \}$ using Martin-Siggia-Rose path integrals. This reads
\be 
{\cal P}(\{ \rho(x,t) \}) \propto \left \langle \int {\cal D}\psi\, e^{\left[\int_0^1 {\rm d}t \int {\rm d}x  N^2 i \psi(x,t) \left(\partial_t \rho
+ \partial_x J \right) \right]} \right \rangle_\xi
\ee
Performing the average over $\xi$ gives the following action (and renaming $- i \psi \to \psi$):
\bea
\label{action_method2}
\nonumber
{\cal S} &= & N^2 \int_0^1 {\rm d}t \int {\rm d}x \left[  \psi \partial_t \rho + f(x,t) \rho \partial_x \psi \right. \\
&-& \left. \frac{\psi}{2N} \partial^2_{xx} \rho 
+ \frac{1}{2} \rho (\partial_x \psi)^2 \right]
\eea
with $f(x,t) = \int {\rm d}y \partial_x V(x-y) \rho(y,t)$.  Taking functional derivatives with respect to $\rho$ and $\psi$ then leads to the following set of equations:
\be
\partial_t \rho = \partial_x (\rho f) + \partial_x (\rho \partial_x \psi) + \frac{1}{2N} \partial^2_{xx} \rho
\ee
and
\bea \nonumber
\partial_t \psi &-& \frac12 (\partial_x \psi)^2 =  f \partial_x \psi - 
\frac{1}{2N} \partial^2_{xx} \psi\\ 
&-& \partial_x \int {\rm d}y \, V(x-y) \rho(y,t) \partial_y \psi(y,t).
\eea
The Euler-Matystin equations are recovered, after a little work, by setting $v(x,t) = - f(x,t) - \partial_x \psi(x,t)$. 
One can finally check \cite{ustocome} that the ${\cal S}$ coincides with $S$ when using the equation of motion satisfied by $\rho, \psi$ and $f$. 
Note that this second method gives rise to additional ``diffusion'' terms, of order $1/N$, which lead to a viscosity term in the velocity equation. 
This second method might therefore be more adapted to search for subleading corrections (in $N^2$) to the action. 

Somewhat surprisingly, Matytsin's formalism has not been exploited to find explicit solutions for $F_{\beta}(A,B)$ in some special cases. One fully solvable case is when
$A$ and $B$ have centered Wigner semicircle spectra \footnote{The case where $\Tr A,B \neq 0$ can be easily treated by noting that the Euler-Matytsin equations are invariant under Galilean transformations, which allows one 
to recover the trivial result from shifting $A,B \to A,B - \mathbb{I}_N N^{-1} \Tr A,B$ in the HCIZ integral, Eq. \ref{HCIZ-def}.}, $\rho_{A}(\nu)=\sqrt{4 \sigma^2_A - \nu^2}/2\pi \sigma_A^2$, and, similarly, for $\rho_B$, 
with a width $\sigma_B$.
One can first note that since trivially $F_{\beta}(A,B)=F_{\beta}(A/z,zB)$, one can always choose $z=\sqrt{\sigma_A/\sigma_B}$ and set $\sigma_A=\sigma_B=\sigma$. The second remark is that the Euler-Matytsin equations 
can be solved by choosing $\rho^2(x,t) = \alpha(t) + \gamma(t) x^2$ and $v(x,t)=b(t) x$, which leads to ordinary differential equations for $\alpha, \gamma$ and $b$.
The final solution is that $\rho(x,t)$ is a Wigner semicircle for all $t$, with a width $\Sigma(t)$ given by
\be
\Sigma^2(t) = \sigma^2 + g t(1-t), \qquad g= \sqrt{1+4 \sigma^4} - 2\sigma^2,
\ee
and $b(t)= (t-1/2)g/\Sigma^2(t)$. Note that $\Sigma^2(t=0)=\Sigma^2(t=1)=\sigma^2$, as it should be. Injecting these expressions into Eqs. (\ref{action_instanton}), (\ref{final}) finally leads to (with $\sigma^2=\sigma_A \sigma_B$)
\be\label{finalW}
F_{2,W}(A,B)= \frac{1}{2}\left[\sqrt{4\sigma^4 + 1} - 1 - \log\left( \frac{1 + \sqrt{4\sigma^4 + 1}}{2} \right) \right].
\ee
For arbitrary matrices $A,B$, the narrow spectra limit (corresponding to $\sigma \to 0$) has been worked out by Collins \cite{collins2003moments}. Specializing
his general result to the case of Wigner matrices, one finds:
\be
\label{leading_contrib_sigma_0}
F_{2,W}(A,B) \underset{\sigma \rightarrow 0}{=} \frac{\sigma^4}{2} - \frac{\sigma^8}{4} + \frac{\sigma^{12}}{3} + \mathcal{O}(\sigma^{16}),
\ee
which coincides with the small $\sigma$ expansion of Eq. (\ref{finalW}). In the opposite limit $\sigma \rightarrow \infty$, we find from Eq. (\ref{finalW}):
\be 
\label{leading_contrib_sigma_infty}
F_{2,W}(A,B) \underset{\sigma \rightarrow \infty}{=} \sigma^2 - \ln(\sigma) - \frac{1}{2} - \frac{\sigma^{-2}}{8} + \frac{\sigma^{-6}}{384} + 
\mathcal{O}\left(\sigma^{-10}\right).
\ee 
Note that Eq. (\ref{finalW}) has a singularity (in the complex plane) for $\sigma^4 = -1/4$. The general analytical properties of $F_\beta$ have attracted a lot 
of attention recently, see \cite{goulden2011monotone} and references therein. 

The limit $\sigma \rightarrow \infty$ can be called the dilute limit and can be studied in full generality, since the solution of the Euler-Matytsin equations can be constructed as 
a power series of $\varepsilon=1/\sigma$, where we define $\sigma^2 \equiv \int {\rm d}x \, x^{2} \rho_{A}(x)$ (we choose here, without loss of 
generality \footnote{see previous footnote.}, $\Tr A = \Tr B = 0$, and rescale the matrices $A,B$ appropriately such that both have the same variance $\sigma^2$). In 
the case where $\rho_A=\rho_B$ but of arbitrary shape (but provided $\rho_A$ vanishes at the edge of the spectrum), 
our final result to order $\varepsilon^6$ reads:
\bea 
\label{dilute_A_eq_B}
& & F_{2}(A,A) \underset{\varepsilon \rightarrow 0}{=} \int {\rm d}x \, x^{2} \rho_{A}(x)  \nonumber \\
&-&  \int {\rm d}x {\rm d}y \, \rho_{A}(x) \rho_{A}(y) \ln|x-y| - \frac{3}{4} - \frac{\pi^2}{6} \int {\rm d}x\,  \rho_{A}^3(x) \nonumber \\ 
&+& \frac{\pi^4}{24} \int {\rm d}x\,  \rho_{A}^3(x) \rho_{A}'(x)^2  + \mathcal{O}(\varepsilon^{10}) 
\eea
which is identical to Eq. (\ref{leading_contrib_sigma_infty}) when $\rho_A$ is a Wigner semicircle, but holds more generally. Note that terms appear in order of 
importance in the above formula. 

The general expression for $\rho_A \neq \rho_B$ is cumbersome and will be given in a longer version of this work \cite{ustocome}. To order $\varepsilon^2$, 
the result  reads:
\bea 
& & F_{2}(A,B) \underset{\varepsilon \rightarrow 0}{=} \int_0^1 {\rm d}p\, X_{A}(p)X_{B}(p) \nonumber \\
&-& \frac12 \int {\rm d}x {\rm d}y \, \rho_{A}(x) \rho_{A}(y) \ln|x-y| \nonumber \\
&-& \frac12 \int {\rm d}x {\rm d}y\,  \rho_{B}(x) \rho_{B}(y) \ln|x-y| - \frac{3}{4} \nonumber \\
&-& \frac{\pi^2}{6} \int_0^1 {\rm d}p \, \rho_{A}[X_A(p)]\, \rho_{B}[X_B(p)] + \mathcal{O}(\varepsilon^{6}) 
\eea
where $X_Z(p)$ is such that $p = \int_{-\infty}^{X} {\rm d}u \rho_Z(u) \in [0,1]$. For $A = B$, one recovers Eq. (\ref{dilute_A_eq_B}) 
by changing variables back from $p$ to $x$, with the Jacobian $dp/dx = \rho_{A}(x)$. The leading term in the above expansion is in fact 
$\int_0^1 {\rm d}p\, X_{A}(p)X_{B}(p)$ and is easy to interpret: it comes from the fact that in the limit $\sigma \to \infty$, HCIZ integrals Eq. (\ref{HCIZ-def}) 
are dominated by the matrix $\Omega$ that diagonalizes $B$ in the diagonal base of $A$ (and the corresponding eigenvalues $\{\lambda\},\{\nu\}$ are ordered). 

The main achievements of this work are twofold: we first rederived the large $N$ asymptotics of HCIZ integrals, first obtained by Matytsin, using Dyson's Brownian motion and the method of instantons. 
We also provided an exact, explicit solution for the case of Wigner matrices, as well as a general expansion method in the dilute limit, when the eigenvalue spectra spread over very wide regions. 
Beyond providing a relatively straightforward and transparent interpretation of Matytsin's method, our
work could provide a valuable starting point to obtain new results, such as the generalization to other ensembles (orthogonal, symplectic, Wishart), as in \cite{guionnet2004}, but also to understand
the structure of subleading (in $N^2$) corrections. Our explicit results in the dilute limit should also be useful for applications, such as, for example, the 
Bayesian estimate of large correlation matrices using empirical data \cite{usinprep}. 

We thank R. Allez, J. Bonart, R. Chicheportiche, A. Guionnet, and J. B. Zuber for useful comments and suggestions.
S.N.M acknowledges support from ANR Grant No. 2011-BS04-013-01 WALKMAT.

\bibliography{LargeN_HCIZ_PRL}

\end{document}